\documentclass[reprint,superscriptaddress,amsmath,amssymb,aps,prl,nofootinbib]{revtex4-2}
\usepackage{soul}
\usepackage{textcomp}  % Additional text symbols
\usepackage{graphicx}  % Include figure files
\usepackage{hyperref}  % Add hypertext capabilities
\usepackage{siunitx}
\usepackage{xspace}
\usepackage{tabularx}
\usepackage[symbol]{footmisc}
\usepackage{IEEEtrantools}
\DeclareSIUnit\gauss{G}
\DeclareSIUnit\rad{rad}
\DeclareSIUnit\dbc{dBc}

\newcommand{\ket}[1]{\ensuremath{\lvert #1 \rangle}\xspace}
\newcommand{\bra}[1]{\ensuremath{\langle #1 \rvert}\xspace}

\newcommand{\figref}[2][]{Fig.\,\ref{#2}#1}

\newcommand{\tripletptwo}{\ensuremath{^3P_{2,m_j=0}}}

\newcommand{\PRL}{Phys. Rev. Lett.}

\newcommand{\PRX}{Phys. Rev. X}
\newcommand{\PRA}{Phys. Rev. A}

\newcommand{\JPhysB}{J. Phys. B}
\newcommand{\RevModPhys}{Rev. Mod. Phys.}
\newcommand{\PhysRevApplied}{Phys. Rev. Applied}
\newcommand{\NatPhys}{Nat. Phys.} %Nature Physics
\newcommand{\NatComm}{Nat. Commun.} %Nature Communications
\newcommand{\PhysRevRes}{Phys. Rev. Res.}
\newcommand{\NewJPhys}{New J. Phys.} %New Journal of Physics

\newcommand{\atomatom}{\SI{715\pm30}{\milli\second}} %g(2) function

\newcommand{\sr}{\text{$^{88}$Sr}}

\makeatletter
\def\maketitle{
	\@author@finish
	\title@column\titleblock@produce
	\suppressfloats[t]}
\makeatother

\begin{document}
	
    \newcommand{\partitle}[1]{\section{#1}}

    \newcommand{\papertitle}{Realization of a fast triple-magic all-optical qutrit in \sr{}}
    %\newcommand{\papertitle}{Fast all-to-all coupling of the strontium clock- and ground-state under triple-magic conditions}
    %Fast, multi-photon, triple-magic (not qutrit?), all-to-all coupling

	\title{\papertitle{}}
	
	\author{Maximilian Ammenwerth}%\thanks{These authors contribute equally to this work.}
	   \affiliation{Max-Planck-Institut f\"{u}r Quantenoptik, 85748 Garching, Germany}
	   \affiliation{Munich Center for Quantum Science and Technology (MCQST), 80799 Munich, Germany}

    \author{Hendrik Timme}%\thanks{These authors contribute equally to this work.}
	   \affiliation{Max-Planck-Institut f\"{u}r Quantenoptik, 85748 Garching, Germany}
	   \affiliation{Munich Center for Quantum Science and Technology (MCQST), 80799 Munich, Germany}

    \author{Flavien Gyger}%\thanks{These authors contribute equally to this work.}
	   \affiliation{Max-Planck-Institut f\"{u}r Quantenoptik, 85748 Garching, Germany}
	   \affiliation{Munich Center for Quantum Science and Technology (MCQST), 80799 Munich, Germany}

    \author{Renhao Tao}%\thanks{These authors contribute equally to this work.}
	   \affiliation{Max-Planck-Institut f\"{u}r Quantenoptik, 85748 Garching, Germany}
	   \affiliation{Munich Center for Quantum Science and Technology (MCQST), 80799 Munich, Germany}
	   \affiliation{Fakultät für Physik, Ludwig-Maximilians-Universit\"{a}t, 80799 Munich, Germany}

    \author{Immanuel Bloch}%\thanks{These authors contribute equally to this work.}
	   \affiliation{Max-Planck-Institut f\"{u}r Quantenoptik, 85748 Garching, Germany}
	   \affiliation{Munich Center for Quantum Science and Technology (MCQST), 80799 Munich, Germany}
	   \affiliation{Fakultät für Physik, Ludwig-Maximilians-Universit\"{a}t, 80799 Munich, Germany}

    \author{Johannes Zeiher}
    \email{johannes.zeiher@mpq.mpg.de}
     \affiliation{Max-Planck-Institut f\"{u}r Quantenoptik, 85748 Garching, Germany}
     \affiliation{Munich Center for Quantum Science and Technology (MCQST), 80799 Munich, Germany}
     \affiliation{Fakultät für Physik, Ludwig-Maximilians-Universit\"{a}t, 80799 Munich, Germany}
	
	\date{\today}
	
	\begin{abstract}
        The optical clock states of alkaline earth and alkaline earth-like atoms are the fundament of state-of-the-art optical atomic clocks.
        An important prerequisite for the operation of optical clocks are magic trapping conditions, where electronic and motional dynamics decouple.
        Here, we identify and experimentally demonstrate simultaneous magic trapping for two clock transitions in \sr{}, realizing so-called triple-magic conditions at a specially chosen magic angle.
        Under these conditions, we operate an all-optical qutrit comprising the ground state $^1S_0$, and the two metastable clock states $^3P_0$ and $^3P_2$.
        We demonstrate fast optical control in an atom array using two- and three-photon couplings to realize high-fidelity manipulation between all qutrit states.
        At the magic angle, we probe the coherence achievable in magic-angle-tuned traps and find atom-atom coherence times between the metastable states as long as \atomatom.
        Our work opens several new directions, including qutrit-based quantum metrology on optical transitions and high-fidelity and high-coherence manipulation on the \sr{} fine-structure qubit.
        
        %At the same time, d
        %Driving the clock transition directly in bosonic isotopes requires large laser intensities and magnetic fields, which typically limits the achievable Rabi frequency on this transition.
        %	
        %Using three phase-coherent light fields we overcome this limitation and demonstrate coherent and fast Rabi coupling using only a weak magnetic field and low laser intensities. 
        %
	%    Furthermore, we identify a triple-magic trapping condition for the ground state $^1S_0$, $^3P_0$ and $^3P_2$ at  $813\,$nm and benchmark the coherence of the corresponding qutrit reaching $T_2$ times of up to several 100ms.
        %
        % Here, we demonstrate a qutrit implemented in the ground-state and two long-lived clock states of \sr. 
		%
		%
		%, in particular in combination with additional constrains, like specific magnetic field angles.
		%
        %Using three- and two-photon couplings we demonstrate all-to-all coupling of the qutrit states.
       % Our work overcomes a number of limitations to using \sr{} for quantum simulation and quantum computing and opens the path towards simultaneous clock interrogation for improved metrological performance. 
	
        \end{abstract}
	%%%%%%%%%%%%%%%%%%%%%%%%%%%%%%%%%%%%%%%%%%%%
	%               Introduction               %
	%%%%%%%%%%%%%%%%%%%%%%%%%%%%%%%%%%%%%%%%%%%%
	%\section{Introduction}
	%
        \maketitle
        
	\begin{figure*}[t!]
		\centering
		\includegraphics[width=\textwidth]{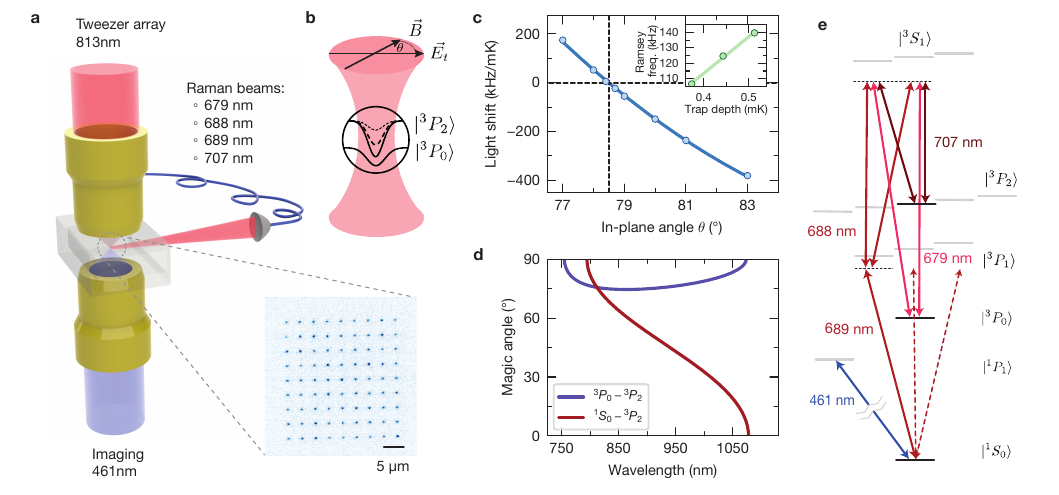}
		\caption{
			\textbf{Realization of triple-magic trapping conditions.} 	\textbf{a} Main features of our experimental setup. We generate tweezer arrays at a wavelength of \SI{813}{\nano\meter} using the top objective and image the atoms onto a camera using the bottom objective.
            Coherent control over the qutrit states is realized with four light fields at $679\,$nm, $688\,$nm, $689\,$nm, and $707\,$nm, which are delivered with common linear polarization from the same optical fiber and focussed onto the tweezer array.
            %For the control of the atomic electronic states, four controllable light fields at wavelength 679 nm, 688 nm, 689 nm and 707 nm are carried in one optical fibre and focused onto the atoms from the side.
			%
			\textbf{b} Trapped in the optical tweezer, the two clock states a priori experience different trapping potential with a strong contribution of the tensor-polarizability. By choosing a suitable angle of the magnetic field with respect to the tweezer polarization we equalize the polarizability as described in the main text.
			\textbf{c} The differential light shift between the clock states obtained from Ramsey spectroscopy is shown as a function of the magnetic field angle. The dependence of the differential light-shift on tweezer depth is shown as a function of trap depth in the inset for a specific magnetic field in-plane angle of \SI{81}{\degree}.
			\textbf{d} The expected magic angle is shown as function of wavelength. At $813\,$nm the $^1S_0$ and $^3P_0$ state are scalar-magic which therefore enables triple-magic conditions.
			\textbf{e} Level diagram of \sr{} and the relevant wavelength used in this work. All light fields share a common linear polarization which is tilted with respect to the quantization axis, resulting in approximately equal projections for the $\pi$ and $\sigma^{\pm}$ contributions. The polarizations and resulting couplings are discussed in the main text and section A of the Appendix. To drive three-photon transitions, we choose a detuning in the vicinity of the ${^1S_0} \leftrightarrow {^3P_{1,m_j=-1}}$ resonance.
	}
	\label{fig:1}
	\end{figure*}

    %https://journals.aps.org/prapplied/abstract/10.1103/PhysRevApplied.19.064024
    %https://journals.aps.org/pra/abstract/10.1103/PhysRevA.97.022115
    The vast majority of quantum information and quantum metrology protocols relies on binary representations of quantum information in terms of qubits.
    For example, efficient interferometry sequences have been devised in quantum metrology that translate phase information into measurable qubit populations~\cite{Hinkley2013,Kessler2014,Young2020}.
    Similarly, universal gate sets are known for qubits in quantum computation~\cite{Bravyi2005,Saffman2016,Saffman2010,Evered2023,Shi2018}, and error correction architectures for qubits are well studied~\cite{Fowler2012,Sahay2023,Knill1997,Wu2022}. 
    In practice, however, most quantum systems feature more states, which, if controlled, can serve as valuable resources~\cite{Roy2023,Shlyakhov2018,Omanakuttan2024}.
    Examples include motional states in trapped atoms or ions or additional internal states that extend the Hilbert space beyond the qubit subspace~\cite{Bohnmann2024,Ringbauer2022}.
    The optical clock states of \sr{} are particularly interesting for  applications in quantum computing~\cite{Schine2022,Madjarov2020} and metrology~\cite{Young2020,Shaw2024}.
    So far, most experiments have used the $^3P_0$ state, coupled to the electronic ground-state $^1S_0$ at a wavelength of \SI{698}{\nano\meter} as an optical qubit.
    However, \sr{} features a much richer structure, including the metastable $^3P_2$ state, which has a predicted natural lifetime exceeding $100\,$s, providing a second clock transition at $671\,$nm when coupled to the ground-state $^1S_0$~\cite{Trautmann2023}.
    In principle, control over the three transitions coupling the metastable states $^3P_0$, \tripletptwo{} and $^1S_0$ thus provides access to an all-optical qutrit, where all states are split by optical frequencies and can be coupled by laser drives.
    Compared to approaches with encodings in motional or hyperfine-structure states, the encoding of a qutrit of states split by optical transition frequencies offers new perspectives, including the tailored motional-state coupling~\cite{Shaw2024}, excellent state-preparation and measurement fidelities and novel readout schemes of qubit states~\cite{Minev2019}.

    Experimentally, beyond the well-studied ${^1S_0}\leftrightarrow{^3P_0}$ transition, coherent manipulation of the $^3P_2$ state was demonstrated~\cite{Klsener2024}.  
    Furthermore, recent work has explored the fine-structure qubit, which involves only the metastable clock states ${^3P_0}\leftrightarrow {\tripletptwo{}}$~\cite{Pucher2024,Unnikrishnan2024}.
    %
    %This qubit allows for fast single-qubit rotations using a Raman coupling based on two-light fields far detuned from an intermediate excited state.
    %
    Here, the atoms were trapped at specific, experimentally fine-tuned magic trapping conditions, where differential light shifts induced by the trapping light vanish and internal and external motional degrees of freedom decouple.
    Such conditions have been demonstrated for various optical transitions in alkaline-earth (like) atoms by tuning the angle between the trap polarization and the quantization axis defined by a magnetic field~\cite{Norcia2018,Ido2003,Yamamoto2016}, or ellipticity of the trap polarization with respect to the quantization axis~\cite{Cooper2018,Klsener2024,Pucher2024}.
    %
    %Preparing this qubit typically requires a population transfer from the ground-state to the $^3P_0$
    %Even at magic trapping conditions, an experimental difficulty in coupling the optical clock states, in particular in bosonic \sr{}, is the requirement for large laser power in combination with strong magnetic fields~\cite{Taichenachev2006}.
    %
    %The magnetic fields are necessary to enable the clock transition by coupling $^3P_1$, which usually limits the achievable Rabi frequency~\cite{Taichenachev2006} to xxx range. 
    %
    %This severe experimental limitation can be mitigated by employing a three-photon coupling scheme~\cite{Hong2005}, which was recently demonstrated for the ${^1S_0}\leftrightarrow{^3P_0}$ transition using a Bose-Einstein condensate~\cite{He2024} and a thermal cloud of strontium atoms~\cite{Carman2024}.
    %

    %%%%%%%%%%%%%%%%%%%%%%%%%%%%%%%%%%%%%%%%%%%%
    %               Here, we show...           %
    %%%%%%%%%%%%%%%%%%%%%%%%%%%%%%%%%%%%%%%%%%%%
    Here, we significantly expand on previous work and implement a highly coherent all-optical qutrit in \sr{} that is realized between both metastable clock states $^3P_0$, \tripletptwo{} and the ground state $^1S_0$ with several novel features:
    \textit{First}, we demonstrate the decoupling of all electronic qutrit states from motional states at magic trapping conditions in an optical tweezer array at $813\,$nm.
    %
    %Our work relies on tuning to simultaneous magic trapping conditions at a trap wavelength of $813\,$nm for the ground state and both clock states.
    %
    %We utilize the tensor polarizability of the $^3P_2$ state to eliminate differential light shifts with respect to $^3P_0$ by applying a magnetic field under a suitable angle. 
    %
    \textit{Second}, at magic conditions, we demonstrate coherent coupling of all involved qutrit states via multi-photon transitions~\cite{Hong2005}, which allow for fast all-to-all Rabi coupling at small magnetic fields and optical powers compared to those typically required for coupling directly on optical clock transitions in \sr{}~\cite{Taichenachev2006}.
    In particular, our work significantly surpasses the coherence demonstrated recently for the ${^1S_0}\leftrightarrow{^3P_0}$ transition using a Bose-Einstein condensate~\cite{He2024} and a thermal cloud of strontium atoms~\cite{Carman2024}.
    %
    %The magnetic fields are necessary to enable the clock transition by coupling $^3P_1$, which usually limits the achievable Rabi frequency~\cite{Taichenachev2006} to xxx range. 
    %
    %This severe experimental limitation can be mitigated by employing a three-photon coupling scheme, which was recently demonstrated for the ${^1S_0}\leftrightarrow{^3P_0}$ transition using a Bose-Einstein condensate~\cite{He2024} and a thermal cloud of strontium atoms~\cite{Carman2024}.
    %three-photon pulse to prepare the $^3P_0$ clock state using single atoms trapped in a tweezer array at a wavelength of \SI{813}{\nano\meter}.
    %
    %To measure the trap-induced differential light shift we carry out Ramsey spectroscopy on the fine-structure states and find a magic angle of \SI{78.5}{\degree} in good agreement with expectations.
    %
    %\textit{Third}, we benchmark the coherence of the qutrit by determining the atom-laser coherence between the two metastable states ${^3P_0}$ and \tripletptwo{} forming the fine-structure qubit in \sr{}.
    \textit{Third}, we benchmark the coherence of the qutrit by characterizing the atom-laser coherence for all three couplings.
    We probe the fundamental limits of the qutrit coherence, which is susceptible to polarization gradients, via a detailed benchmarking of the coherence between the two metastable states ${^3P_0}$ and \tripletptwo{} forming the fine-structure qubit in \sr{}.
    We reach $T_2$ times up to $\tau_{dd}=\SI{345\pm12}{\milli\second}$ under continuous dynamical decoupling, significantly exceeding the previous state of the art on the fine-structure qubit.
    We furthermore find a lower bound of the atom-atom coherence time for the fine-structure qubit as long as $\tau_{at}=\atomatom$, which is on par with coherence times achievable with ground hyperfine states in alkali atoms~\cite{Levine2022,Bluvstein2024}.
    Our demonstration of long coherence times in polarizability-engineered tweezer traps substantiates the scalability of microscopically controllable neutral-atom quantum systems due to its applicability over a broad wavelength range.
    Furthermore, our work opens a new route to atomic state engineering using qutrits.
    %
    %In addition, we demonstrate coherent all-to-all transfer of the qutrit states using a three-photon pulse on the $^1S_0$ to $^3P_2$ transition.
	%
    %

    %%%%%%%%%%%%%%%%%%%%%%%%%%%%%%%%%%%%%%%%%%%%
    %         Describe the main idea           %
    %%%%%%%%%%%%%%%%%%%%%%%%%%%%%%%%%%%%%%%%%%%%
    %\section{Magic angle spectroscopy}
    %An important prerequisite for realizing an all-optical qutrit is the presence of triple-magic trapping conditions for all involved electronic states, where the external motional-state dynamics completely decouples from the internal states.
    %
    In our experiment, we operate a $9\times9$ site tweezer array in the focal plane of a high-resolution objective, see~\figref{fig:1}{a}.
    To realize triple-magic trapping conditions, we choose our tweezer wavelength at \SI{813}{\nano\meter} to provide magic trapping conditions between the ground-state $^1S_0$ and the clock state $^3P_0$~\cite{Takamoto2005,Ludlow2006}.
    In addition, we tune the polarizability of the $^3P_2$ state to the same value, which can be achieved in \sr{} for the \tripletptwo{} state by applying a magnetic field under an angle $\theta$ with respect to the linear tweezer polarization, in analogy to the known strategy for $^3P_1$~\cite{Norcia2018}.
    Due to the absence of total angular momentum $J$ in the ground and $^3P_0$ clock states, their polarizability is unchanged by applying a magnetic field.
    We thus obtain the following expression for the tunable differential polarizability between $^1S_0$, $^3P_0$ and \mbox{\tripletptwo{}}
	\begin{equation}
		\Delta \alpha = 3 \left( \alpha_1 - \alpha_0 \right) \sin^2 \theta + \alpha_0, \label{eqn:diff_light_shift}
	\end{equation}
	with $\alpha_i = \alpha_{^3\!P_0} - \alpha_{^3\!P_2,m_j=i}$ being the polarizability difference between the $J=0$ states and the $m_j=i$ Zeeman state of $^3P_2$.
    Using the known polarizability data for strontium, magic-angle tuning should enable the realization of magic trapping conditions across a wide range of wavelengths, see~\figref{fig:1}{d}.

    Our experimental sequence starts by loading the tweezer array from a two-stage magneto-optical trap, followed by a parity projection step to remove double occupancy in the tweezers.
    Subsequently, we image the occupation of the array by collecting the atomic fluorescence at a wavelength of \SI{461}{\nano\meter} using a second objective and image it onto a \mbox{qCMOS} camera, see~\figref{fig:1}{a}.
    During the imaging pulse, we additionally employ attractive Sisyphus cooling on the ${^1S_0}\leftrightarrow{^3P_1}$ transition at \SI{689}{\nano\meter}, resulting in high-fidelity, low-loss detection of the occupation of individual tweezers~\cite{Covey2019}.
    After this first image, we apply a sideband cooling pulse to cool the atoms in the radial direction, reaching a mean residual excitation number $\bar{n} = 0.14(4)$~\cite{Norcia2018} and correspondingly a motional ground-state occupation of \SI{88\pm 3}{\percent}.
    Subsequently, we apply a magnetic field with a strength of \SI{19}{\gauss} in the horizontal plane at an angle $\theta$ between the linearly polarized tweezer array with electric field vector $\vec{E_t}$ and the magnetic quantization axis $\vec{B}$.
    %, resulting in triple-magic trapping conditions.
    %
    At our typical trap depths of $U_0/k_B\leq \SI{0.5}{\milli\kelvin}$, the chosen field strength is sufficient to define the quantization axis, which is required for magic-angle tuning.
    To couple $^1S_0$ to $^3P_0$ (\tripletptwo{}), we use a combination of three photons at $689\,$nm, $688\,$nm and $679\,$nm ($707\,$nm), coupling via the intermediate states $^3P_1$ with a detuning of $|\Delta_{689}| =2\pi\times 6\,$MHz (with respect to  $^3P_{1,m_j = -1}$) and $^3S_1$ with a detuning of $|\Delta_{679}|= 2\pi\times12\,$GHz ($|\Delta_{707}|= 2\pi\times12\,$GHz), see~\figref{fig:1}{e}.
    We couple the two metastable states ${^3P_0}\leftrightarrow{\tripletptwo{}}$ off-resonantly via $^3S_1$ at $\Delta_{679}$ using two photons at $679\,$nm and $707\,$nm.
    To suppress differential phase noise, all lasers are locked to the same frequency comb and delivered to the experiment using the same polarization maintaining fiber. 
    To locate the position of the magic angle, we first prepare the atoms in the $^3P_0$ clock state using a fast three-photon transfer\cite{He2024,Carman2024}, further described below.
    We then measure the light-shift on the ${^3P_0}\leftrightarrow{\tripletptwo{}}$ transition via Ramsey spectroscopy as a function of trap depth at various angles, see~\figref{fig:1}{c}.
    We find that the differential polarizability vanishes at an angle of $\theta_m=\SI{78.49\pm0.03}{\degree}$, in good agreement with the expected value considering a systematic uncertainty in the relative angle between tweezer polarization and magnetic field axis given by the coil orientation.
    \begin{figure}
		\centering
		\includegraphics{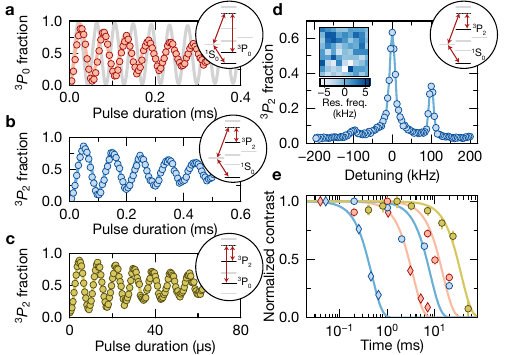}
		\caption{
			\textbf{Three-photon coupling between $^1S_0$ and $^3P_0$, $^3P_2$.} 
			\textbf{a} Rabi oscillations between $^1S_0$ and $^3P_0$. The red line shows simulation results of a simplified two-level model based on the measured in-loop laser phase noise without any free parameter. The gray line shows a noise-free simulation of the system taking all Zeeman-substates and their couplings into account (see Appendix). 
            %The fitted decay rate is in very good agreement with our analytically computed fidelity limit (see Appendix). 
            %
            \textbf{b} We directly drive the ${^1S_0}\leftrightarrow{\tripletptwo{}}$ transition with a three-photon pulse by exchanging the $679\,$nm light field with a corresponding beam at $707\,$nm.
            \textbf{c} Rabi oscillations on the fine-structure qubit under triple-magic conditions. We first initialize the qubit with a three-photon pulse into $^3P_0$ and then drive Rabi oscillations between the fine-structure states.
            \textbf{d} Tweezer-resolved three-photon spectroscopy on the ${^1S_0}\leftrightarrow{\tripletptwo}$ transition, resulting in narrow resonance features with resolved red and blue motional sidebands. Extracting the center frequency for the transition for each tweezer (inset), we compute the tweezer-resolved polarization angle (see Appendix). At a relative detuning of \SI{100}{\kilo\hertz} from the carrier we observe the radial sidebands of our tweezer array.
            \textbf{e} Coherence of the three-photon coupling. We obtain a $T_2^{*}$ time of $\tau_R^{^3P_0}=\SI{3.4\pm0.2}{\milli\second}$ (red diamonds) for the ${^1S_0} \leftrightarrow {^3P_0}$ three-photon coupling and $T_2^{*}$ time of 
			$\tau_R^{^3P_2}=\SI{470\pm18}{\micro\second}$ (blue diamonds) for the $^1S_0 \leftrightarrow {^3P_2}$ qubit, compatible with limitations caused by polarization gradients across the array as discussed in the main text. 
			With a single spin-echo we measure a $T_2$ time of $\tau_{se}^{^3P_0}=\SI{14.8\pm2.2}{\milli\second}$ (red circles) and $\tau_{se}^{^3P_2}=\SI{8.1\pm2.4}{\milli\second}$ (blue circles). For comparison, the $T_2$ time on the fine-structure states under equal conditions reaches $\tau_{se}=\SI{36\pm2}{\milli\second}$ (olive-green circles).
		}
		\label{fig:2}
    \end{figure}

    With the magic condition identified, we proceed to demonstrate fast, coherent, all-optical control of all transitions involved in the qutrit and benchmark the coherence times of all three couplings.
    In particular, we use combinations of three photons to couple from the ground state to both metastable states~\cite{He2024,Carman2024}, and two photons to couple the two metastable states, as recently demonstrated in trapping conditions different from our experiment~\cite{Pucher2024,Unnikrishnan2024}.
    For the qutrit realized here, the combination of multi-photon processes allows for fast all-optical control at relatively modest external magnetic fields compatible with triple-magic angle tuning in experiments.
    %, which dramatically simplifies the experimental requirements to set the magic angle.
    %
    %For the coherent three-photon excitation of the $^1S_0 \rightarrow\,^3P_0$ clock transition, we used three phase-coherent light fields at wavelengths of \SI{689}{\nano\meter}, \SI{688}{\nano\meter} and \SI{679}{\nano\meter} realizing a coupling via the path ${^1S_0} \rightarrow {^3P_1} \rightarrow {^3S_1} \rightarrow {^3P_0}$, see~\figref{fig:1}. 
    %
    %We choose a common linear polarisation for all beams which is tilted approximately \SI{55}{\degree} out of the horizontal plane. 
    %
    The multi-color light field  is impinging onto the array with linear polarization tilted approximately \SI{55}{\degree} out of the horizontal plane, which was chosen to equalize the polarization projection of the $\pi$ and $\sigma^{\pm}$ components with respect to the horizontally set magnetic field and which maximizes the three-photon Rabi frequency in this configuration.
    %, see Appendix for details.
    %
    %To suppress scattering from the intermediate states, we chose a detuning of \SI{6}{\mega\hertz} with respect to the $^3P_{1,m_j = -1}$ state and \SI{12}{\giga\hertz} with respect to the $^3S_1$ state. 
    %
    On three-photon resonance, we observe corresponding Rabi oscillations with an initial contrast (at $\Omega_{{}^{3}\!P_0}\,t = \pi$) reaching up to \SI{88}{\percent} averaged over 118 repetitions of the experiment and the whole tweezer array, see~\figref{fig:2}{a}.
    %, measured by rapidly pushing out the fraction of atoms in $^1S_0$ after a three-photon pulse of variable length. 
    %
    The extracted Rabi frequency of $\Omega_{{}^{3}\!P_0}=2\pi\times\SI{19.16\pm0.02}{\kilo\hertz}$ for less than $10\,$mW total power in all three beams agrees well with analytic expectations and numerical simulations discussed in more detail in the Appendix.
    %At these settings we reached a three-photon Rabi frequency of using around \SI{75}{\micro\watt} at \SI{689}{\nano\meter}, \SI{5}{\milli\watt} at \SI{688}{\nano\meter} and \SI{900}{\micro\watt} at a wavelength of \SI{679}{\nano\meter}.
    %
    By modelling the Rabi oscillations in presence of experimental imperfections, we find that the observed dephasing is almost entirely accounted for by the independently measured laser phase noise~\cite{Ball2016}.
    %, see Appendix. 
    %
    Simulating a noise-free quantum system, we find that the limitations imposed by off-resonant scattering from excited states allow for substantially more Rabi cycles before the system dephases, see \figref{fig:2}{a}.
    In analogy to the coherent coupling to $^3P_0$, we also realize a coherent excitation of the \tripletptwo{} state at the magic angle, by substituting the photon at $679\,$nm with a photon at $707\,$nm, see~\figref{fig:2}{b}.
    Performing high-resolution spectroscopy (shown in~\figref{fig:2}{d}), we find in this case a tweezer-dependent shift of the resonance position, which we attribute to small imperfections in the alignment of the array optics and a corresponding variation of the tweezer polarization angle across the array.
    These observations indicate the critical influence of precise alignment and the challenge to realize scalable tweezer arrays with excellent homogeneity.
    To characterize the coherence of the three-photon coupling, we measure the $T_2^*$ time of the ${^1S_0} \leftrightarrow {^3P_{0}} \,\,({^3P_{2}})$ qubit in a Ramsey sequence, reaching $\tau_R^{^3P_0}=\SI{3.4\pm0.2}{\milli\second}$ $\left(\tau_R^{^3P_2}=\SI{470\pm18}{\micro\second}\right)$. 
    Additionally, we employ a spin-echo sequence to mitigate the effect of polarization gradients and observe enhanced $T_2$ times of $\tau_{se}^{^3P_0}=\SI{14.8\pm2.2}{\milli\second}$ $\left(\tau_{se}^{^3P_2}=\SI{8.1\pm2.4}{\milli\second}\right)$.
    Our measurements, summarized in~\figref{fig:2}{e}, demonstrate coherent all-to-all coupling between the qutrit states.
    We attribute the limited coherence time on the ${^1S_0} \leftrightarrow {^3P_0}$ coupling to laser phase noise, which is presented in the Supplementary material.
    %

    %%%%%%%%%%%%%%%%%%%%%%%%%%%%%%%%%%%%%%%%%%%%
    %          Finestructure qubit             %
    %%%%%%%%%%%%%%%%%%%%%%%%%%%%%%%%%%%%%%%%%%%%
    With fast three-photon control on both ground-to-metastable transitions at hand, we now proceed to benchmark the fundamental coherence limit of the realized qutrit.
    Previous work~\cite{Unnikrishnan2024} has raised concerns as to the achievable coherence times in angle-tuned tweezer potentials due to the presence of spatially dependent differential polarizabilities in tightly focused optical tweezers, which lead to temperature- and trap-depth-dependent spin motion coupling.
    %
    %This limit is determined in magic-angle-tuned traps by polarization gradients, which result in position-dependent differential light shifts and give rise to motion-induced dephasing.
    %
    To characterize the limitations imposed by magic-angle tuning and bypasses the excess phase noise of the \SI{688}{\nano\meter} laser (see Appendix), we characterize the fine-structure (FS) qubit formed by both metastable states coupled on a coherent two-photon transition via the intermediate state $^3S_1$ in triple-magic conditions.
    %
    %Our detailed benchmarking of the FS coherence, described below, characterizes the limitations of the achievable qutrit coherence, imposed by magic angle tuning and bypasses the excess phase noise of the \SI{688}{\nano\meter} laser (see Supplementary material).
    %
    For all measurements described in the following, we initialize the fine-structure qubit via a sideband-resolved three-photon $\pi$-pulse on the ${^1S_0}\leftrightarrow{^3P_0}$ transition, which preserves the motional ground state.
    Detection is similarly performed by controlled de-excitation of the clock state back into the ground state.
    %We prepare atoms in $^3P_0$ using a three-photon $\pi$-pulse with a typical transfer efficiency of about \SI{95}{\percent} and subsequently push out the remaining atoms by rapidly scattering photons on the broad \SI{461}{\nano\meter} transition.
    %
    %To apply the two-photon coupling on the fine-structure qubit, we switch on light fields at \SI{679}{\nano\meter} and \SI{707}{\nano\meter} with a detuning of $\,$GHz from $^3S_1$.
    %
    %To measure the populations of the two fine-structure states we map the $^3P_0$ population back to the ground-state with a second three-photon $\pi$-pulse and apply another pushout pulse before repumping and imaging.
    %
	\begin{figure}
		\centering
		\includegraphics{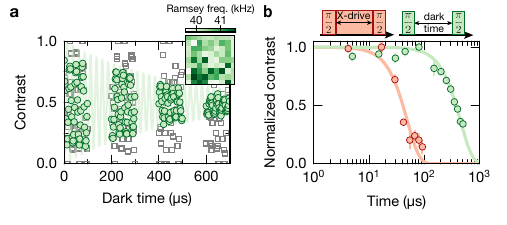}
        \caption{
			\textbf{Fine-structure qubit under triple-magic conditions.} 
			\textbf{a} Tweezer-resolved Ramsey measurement on the fine-structure qubit. The tweezer averaged contrast (green circles) rapidly drops while individual tweezers (gray squares) show a high contrast. The tweezer-dependent Ramsey frequency (inset) is caused by a spatially varying polarization across the tweezer-array (see Appendix). Note that the single tweezer contrast decays on substantially longer timescales.
			\textbf{b} Comparison of the FS Ramsey measurement (green) with a spin-lock measurement (red). For both measurements, we apply two $\pi/2$-pulses along the Y-axis separated by a waiting time in the dark (green) or under a continuous Rabi drive along the X-axis (red). 
            The contrast of the spin-lock measurement decays significantly faster on a time scale of $\tau_{sl}=\SI{49\pm4}{\micro\second}$ in agreement with the contrast decay in the Rabi oscillations, indicating limitations due to laser phase noise. 
            For comparison, the $T_2^{*}$ measurement on the fine-structure qubit shows a tweezer-averaged contrast decay with an extracted time of $\tau_R=\SI{463\pm7}{\micro\second}$. 
			%
            %\textbf{d} Pulse sequence used to carry out Ramsey spectroscopy on the fine-structure qubit.
		}
		\label{fig:3}
	\end{figure}	
    %For a more accurate characterisation of the differential light shift we use a Ramsey type sequence. 
    %
    %After coherently preparing atoms in $^3P_0$ we apply a fine-structure $\pi/2$ pulse followed by a variable waiting time and an additional $\pi/2$ pulse, see \figref{fig:2}{e}. 
	%
    %As function of waiting time we observe that the population oscillates between the clock-states at the Ramsey frequency, see \figref{fig:2}{b}. 
    %
    %The Ramsey frequency is given by the differential light shift which has a contribution from the trapping light and additionally from differential light shifts induced by the Raman pair. 
	%
	%We are interested in the trap-induced differential light shift and obtain the Ramsey frequency at different trap depth, see \figref{fig:2}{c}. 
    %
    %Projected to zero trap depth we obtain an offset which is given by the Raman beam induced differential light shifts that can be eliminated upon choosing a suitable power ratio of the \SI{679}{\nano\meter} and \SI{707}{\nano\meter} beam. 
    %
    %From the slope of the Ramsey frequency as function of trap depth we extract the trap-induced differential light shift shown in \figref{fig:2}{d} as function of the magnetic field angle. 
    %
    %The fit is given by Eq.\,\eqref{eqn:diff_light_shift} in excellent agreement with the measurement.
	%
	%\section{Fine-structure qubit under triple-magic conditions}
    %Under triple-magic trapping conditions we benchmark the fine-structure (FS) qubit
    Driving Rabi oscillations between $^3P_0$ and \tripletptwo{}, we extract a Rabi frequency of $\Omega_{FS} = 2\pi\times\SI{117}{\kilo\hertz}$ using only $\sim\,$\SI{900}{\micro\watt} of power at \SI{707}{\nano\meter} and $\sim\,$\SI{850}{\micro\watt} of power at \SI{679}{\nano\meter}, see~\figref{fig:2}{c}. 
    Here, the polarization setting  that optimizes the coherent three-photon preparation is suboptimal,
    %, since only the $\pi$-projection contributes to the two-photon coupling
    effectively lowering the achievable two-photon coupling on the FS qubit.
    %by a factor of three
    %, see Appendix for a more detailed discussion.
    %
    However, we chose to use this setting for experimental convenience of not having to change the polarization between manipulation of the clock transitions and the FS transition.
    The FS qubit Rabi oscillations exhibit damping with a
    $1/e$ time of \SI{50\pm1}{\micro\second},
    %a rate of about $\gamma = \SI{19.9\pm0.6}{\per\milli\second}$,
    %1/e exp fit decay times instead of rates are 50(1)µs
    again compatible with estimates computed from the independently measured phase noise.
    %, see Appendix.
    %
    To further substantiate that phase noise dominates the decay dynamics of the Rabi oscillations, we perform a spin lock measurement by first applying a $\pi$/2-pulse around the Y-axis, which aligns the state along the X-axis in the state $\ket{+} = \left(\ket{^3P_0} + \ket{\tripletptwo{}}\right)/\sqrt{2}$ followed by a variable-time Rabi drive about the X-axis~\cite{Bodey2019}.
    In the absence of noise, the X-drive is aligned and should stabilize the state $\ket{+}$.
    After a second $\pi/2$-pulse of variable phase, we perform a projective measurement of the populations in $^3P_0$ and \tripletptwo{} and extract the contrast of the resulting Ramsey fringe.
    With this sequence, we observe a fast drop of the contrast with a time constant $\tau_{sl} = \SI{49\pm4}{\micro\second}$ in presence of the X-drive, which indicates that currently excess phase noise at the Rabi drive is a major source of decoherence for our current lasersystem~\cite{Bishof2013} and limits the observed contrast of our Rabi oscillations~\cite{Day2022,Nakav2023}. 
    \begin{figure}[t!]
		\centering
		\includegraphics{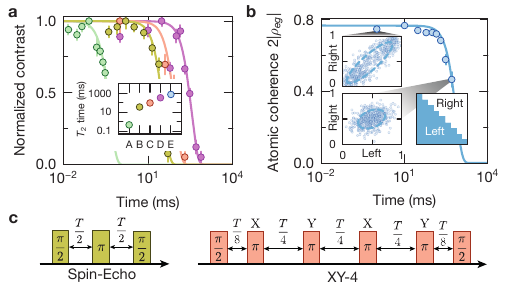}
        \caption{
            \textbf{Coherence of the fine-structure qubit.} \textbf{a} Comparison of the fine-structure qubit $T_2^{*}$ time and $T_2$ time measured with a XY-4 dynamical decoupling sequence. At a trap depth of \SI{46}{\micro\kelvin} we obtain a $T_2^{*}$ time of $\tau_R=\SI{463\pm7}{\micro\second}$ (light-green, inset A) and $T_2$ time of $\tau_{se}=\SI{36\pm2}{\milli\second}$ with a single spin-echo (olive-green, inset B). Using a XY-4 dynamical decoupling sequence the $T_2$ time becomes $\tau_{dd^{\prime}}=\SI{97\pm6}{\milli\second}$ (orange, inset C) and lowering the tweezer depth to about \SI{5}{\micro\kelvin}, we boost the $T_2$ time to $\tau_{dd}=\SI{346\pm37}{\milli\second}$ (purple, inset D) setting a limit on the atom-laser coherence. The inset shows a summary of the measured coherence times including the atom-atom coherence time (blue, inset E).
            \textbf{b} To benchmark the atom-atom coherence we average a $g^{(2)}$ correlation function across subsets of our tweezer array to extract a lower bound for the atomic coherence (see Appendix) and obtain a time constant of \atomatom. The correlations between the left and right half of the tweezer array are shown in the insets.
            \textbf{c} Pulse sequences used to measure the $T_2$ times using spin-echo and XY-4 dynamical decoupling.
        }
		\label{fig:4}
	\end{figure}
    Here, the time constant $\tau_{sl}$ is obtained by fitting the resulting contrast with a function of the form $C(t)\propto\exp\left(- (t/\tau_{sl}) ^2 \right)$.
    %and reaches \SI{463\pm7}{\micro\second} in the absence of the $X$-drive.
    %Subsequently, we turn on a X-drive for variable duration which "locks" the population before we apply a second $\pi$/2 pulse. 
    % our polarisation choice of approximately \SI{55}{\degree} with respect to the quantisation axis (common for all 4 colors) is suboptimal in this case, since only the 
	%
	%While this polarisation maximizes the three-photon Rabi frequency (under the contrain of common linear polarisation) it lowers the fine-structure Rabi frequency by a factor of 3. 
    %
    %See Appendix \ref{app:three_photon} for a more detailed discussion of polarizations. 
    %
    %In the future this compromise can be improved using tailored waveplates or delivering the light with different fibers which we avoid here to suppress differential phase noise.
    To further analyze the performance of the FS-qubit, we also study its coherence in the absence of any driving light.
    %
    %Previous work~\cite{Unnikrishnan2024} has raised concerns as to the achievable coherence times on the fine-structure qubit due to the presence of spatially dependent differential polarizabilities in tightly focussed optical tweezers, which lead to temperature- and trap-depth dependent spin motion coupling.
    %
    First, we perform a standard Ramsey sequence with two $\pi/2$-pulses separated by a variable dark-time and obtain a $T_2^{*}$ time of $\tau_R=\SI{463\pm7}{\micro\second}$. 
    A tweezer-resolved analysis of the Ramsey oscillation frequency reveals that this rapid dephasing is dominated by inhomogeneities across the tweezer array, see~\figref{fig:3}{a}. 
    %
    %At time-scales of several hundred microseconds we observe Ramsey oscillations with high-contrast for each individual tweezer which however oscillate out-of-phase with respect to each other and thereby limit the tweezer-averaged $T_2^{*}$ time. 
    %
    Such inhomogeneous shifts can be mitigated using dynamical decoupling techniques.
    Employing a standard XY-4 dynamical decoupling sequence (see \figref{fig:4}{c}), we find a strongly boosted $T_2$ time of $\tau_{dd}=\SI{345\pm12}{\milli\second}$, exceeding previous measurements in angle- or polarization-tuned magic optical traps by at least an order of magnitude~\cite{Pucher2024,Unnikrishnan2024}.
    We further study the limits to the FS-qubit coherence by analyzing correlations between the qubit states measured within subsystems of the tweezer array, see~\figref{fig:4}.
    This measurement reveals the achievable atom-atom coherence, which can exceed the laser-atom coherence significantly~\cite{Young2020}.
    We extract a decay time of $\tau_{at}=\atomatom$ from a fit to our data (see Appendix), which can be interpreted as a lower bound to the atom-atom coherence achievable for the FS qubit in \sr{} and on par with coherence times achieved e.g., in hyperfine states of alkali atoms~\cite{Bluvstein2024}.
    We expect that, with higher quality Rabi oscillations and more dynamical decoupling pulses, the coherence time can be boosted further.
    Our demonstration of the all-to-all connectivity of the qutrit states under common conditions opens up new possibilities in quantum computing~\cite{Majumdar2018}, quantum simulation, e.g. for the simulation of spin-1 systems~\cite{Mgerle2025,Edmunds2024,Kumaran2024}, and metrology~\cite{Ishiyama2023} with alkaline-earth atoms. 
    %
    %Using phase-coherent laser we drive two- and three-photon transitions which enable all-to-all coupling at high Rabi frequency using only weak magnetic fields and low laser intensities.
	%
    %We benchmark the coherence of the FS qubit under triple-magic trapping conditions and demonstrate that dynamical decoupling techniques are well suited to overcome dephasing rates given by inhomogeneous tweezer polarisation reaching atom-laser coherence times of serveral hundred \si{\milli\second}.
	%
    %In addition, we demonstrate for the first time the direct three-photon coupling between the ground-state and the $^3P_2$ clock-state. 
	%
    Overcoming the technical limitations associated with laser noise opens the route to quantum computing and quantum simulation architectures involving fast transfers between the highly coherent strontium clock qubit~\cite{Young2020} for storage and the fast FS-qubit for operations.
    Realizing common trapping conditions for all qutrit states, comprising the clock states and the ground state, separated by optical frequencies, enables highly configurable interactions for quantum simulation experiments~\cite{vanBijnen2015}.
    State-resolved readout of the qutrit population is feasible by imaging the ground-state population using fast, destructive fluorescence detection on the broad ${^1S_0} \leftrightarrow {^1P_1}$ transition~\cite{Su2025} combined with fast three-photon pulses to sequentially map the clock state populations to the ground state.
    Compared to the usually employed single-photon couplings, the three-photon coupling has the distinct advantage of operating at only low to moderate magnetic fields with high Rabi couplings at small required optical powers, dramatically improving the scaling perspectives of this approach (see Appendix for details).
    Tuning the two-photon coupling from the clock states on resonance with the $^3P_1$ state, which subsequently decays to the ground state, can serve as a building block for recently proposed measurement-free quantum error correction~\cite{Heussen2024}, or highly controllable dissipative steps required for digital quantum simulation of open systems~\cite{Barreiro2011}.
    A re-initialization back to the clock states is subsequently possible via a fast three-photon pulse, exploiting the all-to-all connectivity of the qutrit.
    Going beyond the qutrit benchmarking, our promising results on the achievable coherence in angle-tuned magic traps open interesting perspectives for using the ${^1S_0}\leftrightarrow  {\tripletptwo{}}$ qubit, which can conveniently be tuned to magic conditions across a broad wavelength range, including IR wavelengths suitable for high-power Ytterbium-doped fiber amplifiers.
    Sideband-resolved three-photon pulses on this transition and a dissipative state reset via the $^3P_1$ state enable direct sideband cooling, providing a novel route to ultralow-entropy states in assembled Hubbard systems.
    %
    %Reaching long coherence times with dynamical decoupling, furthermore has applications for magnetic field sensing where the decoupling pulses can be tailored to optimize the sensitivity of the sensor~\cite{Pham2012}.
    %
    Finally, our demonstrated atom-atom coherence time involving the $^3P_2$ state opens promising perspectives for a variety of clock transitions in atoms trapped in non-scalar magic optical traps~\cite{Ishiyama2023}. 
    Thus, our work sets the stage for quantum metrology applications involving at least two clock transitions, which have been discussed in the context of uncovering slow temporal changes in fundamental constants~\cite{Safronova2018}.
    To further suppress the effect of locally varying polarization, we envision to trap atom arrays in optical lattices instead of tweezer arrays~\cite{Young2022,Buob2024,Tao2024,Gyger2024}.
    Furthermore, single- and two-qubit gates on the FS-qubit will benefit from erasure conversion based on the fluorescence detection of ground-state population~\cite{Wu2022,Su2025,Ma2023}.
    Using combinations of single-qutrit rotations and Rydberg-mediated two-qubit gates, coupling either of the $^3P_J$ states to a Rydberg state, we envision creating and certifying three-dimensional entanglement~\cite{Friis2018}.
	%
    %Our results open up new possibilities in quantum computing and metrology where the coherence can be transfered between different states optimized for specific applications exploiting the fast coupling of the fine-structure states~\cite{Pucher2024} and the long coherence of the scalar-magic $^3P_0$ clock state~\cite{Young2020}.
    
	\begin{acknowledgments}
		% \section*{Acknowledgments}
		We thank Ria Rosenauer and Kevin Mours for help with the \SI{688}{\nano\meter} laser setup and Eran Reches and Lorenzo Festa for the code to estimate phase noise induced limitations.
		We acknowledge funding by the Max Planck Society (MPG) the Deutsche Forschungsgemeinschaft (DFG, German Research Foundation) under Germany's Excellence Strategy--EXC-2111--390814868, from the Munich Quantum Valley initiative as part of the High-Tech Agenda Plus of the Bavarian State Government, and from the BMBF through the programs MUNIQC-Atoms and MAQCS.
		This publication has also received funding under Horizon Europe programme HORIZON-CL4-2022-QUANTUM-02-SGA via the project 101113690 (PASQuanS2.1).
		J.Z. acknowledges support from the BMBF through the program “Quantum technologies - from basic research to market” (SNAQC, Grant No. 13N16265).
		H.T., M.A. and R.T. acknowledge funding from the International Max Planck Research School (IMPRS) for Quantum Science and Technology. M.A acknowledges support through a fellowship from the Hector Fellow Academy.
		F.G. acknowledges funding from the Swiss National Fonds (Fund Nr. P500PT\textunderscore203162).
		
	\end{acknowledgments}
	
    \appendix
    \section{A. Three-photon coupling}
\label{app:three_photon}
This Appendix provides a brief analytic description of the three-photon coupling including a fidelity estimation obtained after adiabatic elimination of the intermediate excited states. 
Following Ref.~\cite{Hong2005}, we consider a 4-level system and neglect the Zeeman-substructure. In a suitable rotating frame the Hamiltonian is given by
\begin{IEEEeqnarray*}{rCl}
    H/\hbar &=& - \Delta_{\mathrm{689}} \ket{^3P_1}\bra{^3P_1} - (\Delta_{\mathrm{689}} + \Delta_{\mathrm{688}}) \ket{^3S_1}\bra{^3S_1} \\
    &&- (\Delta_{\mathrm{689}} + \Delta_{\mathrm{688}} -\Delta_{\mathrm{679}}) \ket{^3P_0}\bra{^3P_0} \\
    && + \left( \frac{\Omega_{\mathrm{689}}}{2} \ket{^1S_0}\bra{^3P_1} + \frac{\Omega_{\mathrm{688}}}{2} \ket{^3P_1}\bra{^3S_1} \right. \\
    && + \left. \frac{\Omega_{\mathrm{679}}}{2} \ket{^3S_1}\bra{^3P_0} + h.c. \right).
\end{IEEEeqnarray*}
Here, $\Delta_i$ denotes the detuning from the respective resonance.
For example, $\Delta_{\mathrm{689}}$ denotes the detuning from the ${^1S_0}\leftrightarrow{^3P_{1,m_J=-1}}$ transition and $\Delta_{\mathrm{688}}$($\Delta_{\mathrm{679}}$) denote the detuning from the ${^3P_{1,m_J=-1}} (^3P_0)\leftrightarrow{^3S_{1,m_J=-1}}$ resonance. 
The Rabi frequencies $\Omega_i$ account for the coherent coupling of the multi-chromatic light field and are computed according to Eq.~\ref{eqn:rabi_freq}.
In addition to the unitary coupling described by the Hamiltonian, both intermediate excited states decay due to their finite lifetime which is described by Lindblad decay operators: $L_{ij} = \sqrt{\Gamma_{ij}} \ket{j}\bra{i}$ for $i = {^3P_1}, {^3S_1}$ and $j = {^1S_0}, {^3P_0}$.
We consider decay from $^3P_1$ back to $^1S_0$ and decay from $^3S_1$ to $^3P_0$ and $^3P_1$ (which subsequently decays further to $^1S_0$).
This is particularly relevant for the $^3S_1$-state which has a lifetime of \SI{13.9\pm0.1}{\nano\second} and consequently a decay rate comparable to the Rabi frequencies on the individual legs~\cite{Heinz2020}. 
A large detuning with respect to the $^3S_1$-state is therefore required to overcome limitations from off-resonant single-photon scattering.
Under the combined effect of coherent driving and dissipation the dynamics of our model are described by a Lindblad master equation for a density matrix $\rho$. 
We obtain the effective dynamics upon adiabatic elimination of the two intermediate excited states following an effective operator formalism~\cite{Reiter2012}. 
This provides us with the coherent coupling rates as well as dissipation rates within the ground-state subspace. 
The effective Hamiltonian for the remaining two-level system contains the three-photon coupling and the single-beam light shifts
\begin{IEEEeqnarray*}{rCl}
    H_{\mathrm{eff}}/\hbar &=& - (\Delta_{\mathrm{689}} + \Delta_{\mathrm{688}} -\Delta_{\mathrm{679}}) \ket{^3P_0}\bra{^3P_0} \\
    && + \left( \frac{\Omega_3}{2}\ket{^1S_0}\bra{^3P_0}  + h.c. \right) \\
    && + \frac{1}{4} \frac{\lvert\Omega_{\mathrm{689}}\lvert^2}{\Delta_{\mathrm{689}}} \ket{^1S_0}\bra{^1S_0} + \frac{1}{4} \frac{\lvert\Omega_{\mathrm{679}}\lvert^2}{\Delta_{\mathrm{679}}} \ket{^3P_0}\bra{^3P_0},
\end{IEEEeqnarray*}
using the further approximated three-photon Rabi frequency
\begin{IEEEeqnarray*}{rCl}
    \Omega_3 &=& \frac{\Omega_{\mathrm{689}} \Omega_{\mathrm{688}} \Omega_{\mathrm{679}}}{4 \Delta_{\mathrm{689}} \Delta_{679}}.
\end{IEEEeqnarray*}
In addition, the effective operator formalism provides effective decay rates between the $^1S_0$ and $^3P_0$ states which describe depolarization and dephasing. 
The two most important contributions describe dephasing due to off-resonant scattering, which we combine to one effective dephasing rate
\begin{IEEEeqnarray*}{rCl}
    \Gamma_{\mathrm{eff}} &=& \Gamma_{3P1} \frac{\Omega_{\mathrm{689}}^2}{4\Delta_{\mathrm{689}}^2} + \Gamma_{3S1} \frac{\Omega_{\mathrm{679}}^2}{4\Delta_{\mathrm{679}}^2}
\end{IEEEeqnarray*}
Additional contributions to dephasing and depolarization are suppressed in comparison to these dominant contributions within our 4-level approximation.
We are interested in finding the optimal parameter set in the 5-dimensional parameter space (one detuning is fixed to stay on three-photon resonance) to maximize the fidelity.
We find a simple analytic expression for the infidelity under the assumption that the fidelity is optimal if the scattering rates from both excited states are equal.
Under this assumption, the infidelity estimate $\epsilon$ depends only on the Rabi frequency of the \SI{688}{\nano\meter} beam,
\begin{IEEEeqnarray*}{rCl}
    \epsilon &=& \frac{\Gamma_{\mathrm{eff}}}{\Omega_3} = 2 \sqrt{\frac{\Gamma_{3P1}}{\Gamma_{3S1}}} \frac{\Gamma_{3S1}}{\Omega_{\mathrm{688}}} \approx \frac{1}{19.4} \frac{\Gamma_{3S1}}{\Omega_{\mathrm{688}}}.
\end{IEEEeqnarray*}
We confirm that this infidelity limit is indeed approximately reached throughout a large volume of the parameter space via numerical simulations of the 4-level system without adiabatic approximation. 
These simulations are executed on three-photon resonance given by the constraint
\begin{IEEEeqnarray*}{rCl}
    \Delta_{\mathrm{679}} &=& \Delta_{\mathrm{688}} + \Delta_{\mathrm{689}} - \frac{1}{4} \left( \frac{\Omega_{\mathrm{679}}^2}{\Delta_{\mathrm{679}}} - \frac{\Omega_{\mathrm{689}}^2}{\Delta_{\mathrm{689}}}\right).
\end{IEEEeqnarray*}
Since all laser beams participating in the three-photon coupling are delivered from the same optical fiber, they exhibit an approximately equal beam waist at the atomic position.
This allows to cancel the three-photon probe shifts using settings for which the differential light shifts vanish:
\begin{IEEEeqnarray*}{rCl}
    \frac{\Omega_{\mathrm{679}}^2}{\Delta_{\mathrm{679}}} &=& \frac{\Omega_{\mathrm{689}}^2}{\Delta_{\mathrm{689}}}
\end{IEEEeqnarray*}
Driving the three-photon coupling requires a polarization with $\pi$ as well as $\sigma^{\pm}$ projection for the individual colors.
To maximize the three-photon Rabi frequency in our collinear setup with common polarization, we choose an angle of approximately \SI{55}{\degree} with respect to the horizontal plane.
This configuration provides an equal polarization projection factor of $\frac{1}{\sqrt{3}}$ to each component, which maximizes the three-photon Rabi frequency under the constraint of common polarization.
%
%Note that this constraint lowers the fine-structure Rabi frequency by a factor of 3 in comparison to ideal polarization projections while leaving the scattering rates approximately unaffected. 
%
For the numerical simulation shown in Fig.~2a we generalize the Hamiltonian to include all dipole-allowed couplings and compute the corresponding Rabi frequency between states $\ket{J_0,m_0}$ and $\ket{J_1,m_1}$ according to
\begin{IEEEeqnarray}{rCl}
    \Omega_{i} &=& p_q  \cdot E_i/ \hbar \cdot D_i \cdot \begin{pmatrix}
        J_0 & 1 & J_1 \\
        m_0 & q & -m_1
    \end{pmatrix}\nonumber \\
    && \times \sqrt{2J_1 + 1} \cdot (-1)^{J_0 +
        J_1 + J_{>}-m_1}. \label{eqn:rabi_freq}
\end{IEEEeqnarray}
Here, $p_q$ is a polarization projection factor, $q=-1,0,1$ labels the polarization $\sigma^-,\,\pi,\,\sigma^+$ of the light field ($i=679,688,689$) and $J_>$ denotes the larger value of $J_0, J_1$~\cite{King2008}. The Rabi frequency depends on the electric field strength $E=\sqrt{4P/(\pi w_0 c \epsilon_0)}$ for given power $P$ and beam waist $w_0$ and the reduced dipole matrix element $D=\sqrt{3\epsilon_0 \hbar \lambda_0^3 \Gamma / \left(8\pi^2 \right)}$ on the transition with the inverse lifetime $\Gamma$ and the transition wavelength $\lambda_0$. For our multi-level extension we also consider decay from $^3S_1$ into $^3P_2$.

\paragraph{Scalability of the three-photon coupling}
We experimentally demonstrate a three-photon Rabi frequency of \SI{19.16\pm0.02}{\kilo\hertz} using a power of $P_{\mathrm{689}} = \SI{75}{\micro\watt}$, $P_{\mathrm{688}} = \SI{8.2}{\milli\watt}$ and $P_{\mathrm{679}} = \SI{850}{\micro\watt}$ focussed to an elliptical focus with horizontal (vertical) waist of approximately \SI{240}{\micro\meter} (\SI{90}{\micro\meter}) at detunings of $\Delta_{\mathrm{689}} = \SI{6}{\mega\hertz}$ and $\Delta_{\mathrm{688,679}} \approx \SI{12}{\giga\hertz}$, see Fig.~2 of the main text. 
To estimate the feasibility of drastically enhanced Rabi frequency, approaching the MHz-regime, we consider scaling up the power of the \SI{688}{\nano\meter} and \SI{679}{\nano\meter} beams by a common ratio $\beta$. 
In this case the individual Rabi frequencies $\Omega_{\mathrm{688/679}}$ are enhanced by a factor $\sqrt{\beta}$ implying that the three-photon Rabi frequency scales up linearly $\Omega_3 \propto \beta$. 
Since the scattering rate depends as well linearly on $\beta$, the scattering-induced infidelity per $\pi$-pulse is unaffected. 
At our given beam waist and detunings, it is thus feasible to realize a Rabi frequency of \SI{1}{\mega\hertz} using approximately \SI{410}{\milli\watt} at \SI{688}{\nano\meter}, \SI{43}{\milli\watt} at \SI{679}{\nano\meter} while maintaining the power of \SI{75}{\micro\watt} at \SI{689}{\nano\meter}, which demonstrates the scaling potential of the three-photon coupling. 
Further Rabi frequency boosts are conceivable upon raising the power of the \SI{689}{\nano\meter} beam.
However, compensating for the enhanced off-resonant scattering in this case requires detuning the beam further, which gives rise to (destructive) interference of excitation paths mediated by different $^3P_1$ Zeeman levels.
To reach a Rabi frequency of approximately \SI{1}{\mega\hertz} at equal beam waists, the direct single-photon coupling would require a prohibitively high optical power of \SI{1}{\kilo\watt} at a magnetic field strength of \SI{1000}{\gauss}.
%%%%%%%%%%%%%%%%%%%%%%%%%%%%%%
%%% Polarization gradients %%%
%%%%%%%%%%%%%%%%%%%%%%%%%%%%%%
\section{B. Polarization gradients}
\label{app:polarisation_gradients}
\begin{figure}[t!]
    \centering
    \includegraphics{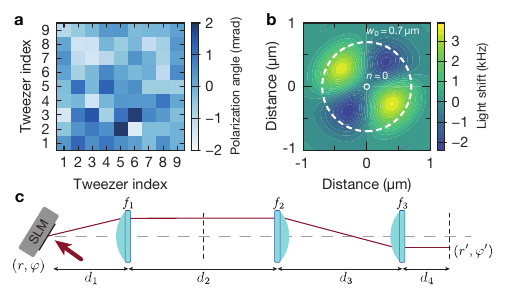}
    \caption{\textbf{Polarization gradients}.
    \textbf{a} Tweezer-resolved polarization angle computed from the tweezer-resolved ${^1S_0}\leftrightarrow{^3P_2}$ spectroscopy.
    \textbf{b} Due to the high numerical aperture used to focus down the tweezer array, the traps exhibit a spatial polarization gradient. We numerically compute the local polarization and resulting differential energy shift shown here for a central trap depth of $U_0/k_B=\SI{46}{\micro\kelvin}$ and a magnetic field strength of \SI{19}{\gauss}. The dashed white ring indicates the waist of our tweezers. The inner solid ring indicates the ground-state root-mean-square value $\sqrt{\langle \hat{x}^2_{n=0} \rangle}$.
    \textbf{c} Simplified sketch of the SLM setup used to estimate the deviation from a 4f-system and its effect on the polarization gradients across the array.} 
    \label{fig:appendix_1}
\end{figure}
The ${^1S_0}\leftrightarrow {^3P_2}$ three-photon resonance shown in Fig.~2d in the main text exhibits a tweezer-dependent resonance shift which is caused by polarization gradients across the tweezer array. 
This shift translates to a tweezer-dependent polarization angle that is consistent with a residual axial misalignment of the SLM setup from a 4f-system.
The tweezer-dependent resonance shift $\Delta f_{\mathrm{res}}$ at a given trap depth $U_0$ is converted to a polarization angle correction $\Delta \theta = \frac{\Delta f_{\mathrm{res}}}{U_0} \frac{1}{\nu}$ using the measured slope $\nu = \SI[per-mode=symbol]{6.03}{\kilo\hertz\per\milli\kelvin\per\milli\rad}/k_B$ (see Fig.~1c in the main text) yielding an angular spread of about $\Delta \theta = \pm\SI{1.6}{\milli\rad}$ across the tweezer array, see \figref{fig:appendix_1}{a}.
%
%We attribute this to a small deviation of our SLM setup from a 4f-configuration.
%
For a displacement $\delta$ of the SLM with respect to a 4f-configuration (see \figref{fig:appendix_1}{c}) the tweezer angle in the atomic plane for a centered input beam ($r=0$) is given by:
\begin{IEEEeqnarray}{rCl}
    \varphi^{\prime} &=& \frac{M \varphi}{f_3} \, \delta \label{eqn:slm_displacement}
\end{IEEEeqnarray}
For our magnification of $M=\frac{f_2}{f_1}=3$, the focal length of $f_3 = \SI{24}{\milli\meter}$ of the microscope objective and a typical angle distribution of about $\varphi(r_{\mathrm{int}}) = \frac{r_{\mathrm{int}}}{f_1} =\pm \SI{2.5}{\milli\rad}$ at the SLM, this indicates a plausible displacement of $\delta \approx \SI{5.3}{\milli\meter}$, given the focal length of $f_1 = \SI{250}{\milli\meter}$ 
and a spatial extent of our tweezer array in the intermediate imaging plane of approximately $r_{\mathrm{int}} =\pm \SI{625}{\micro\meter}$.
Eq.~\ref{eqn:slm_displacement} is derived from a simple ABCD-matrix calculation taking into account the free propagation over distances $d_i$ and the lenses with focal length $f_i$, shown in \figref{fig:appendix_1}{c}.
We find that $\varphi^{\prime}$ is most sensitive to an error $\delta$ on the propagation between the SLM and the first lens considering $d_1 = f_1 + \delta$.
The displacement gives rise to a tweezer-dependent polarization tilt out of the horizontal plane, which consequently contributes to a tweezer-dependent light shift that limits the $T_2^{*}$ time due to inhomogeneous broadening.
In addition to polarization gradients across the tweezer array the local polarization within each tweezers is modulated depending on the displacement with respect to the tweezer center because of the high numerical aperture of $0.65$ of our microscope objective~\cite{Thompson2013}.
The local polarization is computed by numerically solving the vector Debye integrals~\cite{Richards1959} and subsequently translated to energy shifts by numerically computing the eigenstates given the local polarization and trap depth and the globally applied magnetic field~\cite{Cooper2018,Unnikrishnan2024}.
We find that the effect of polarization gradients within the tweezers is irrelevant for $T_2$ times on the order of several hundred milliseconds at our trap depth of $U_0/k_B=\SI{5}{\micro\kelvin}$ even in the absence of axial ground-state cooling.
%

%%%%%%%%%%%%%%%%%%%
%%% Phase noise %%%
%%%%%%%%%%%%%%%%%%%

\section{C. Influence of laser phase noise}
\label{app:phase_noise}\begin{figure}[t!]
    \centering
    \includegraphics{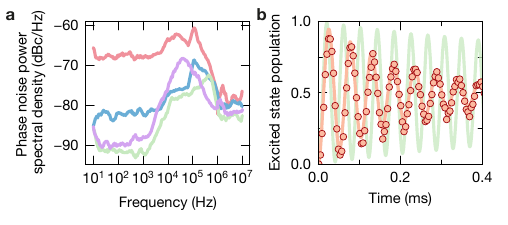}
    \caption{\textbf{Laser phase noise}.
    \textbf{a} Measured phase noise power spectral density obtained from the in-loop beat signal between a common frequency comb and the \SI{688}{\nano\meter} (red), \SI{689}{\nano\meter} (blue), \SI{707}{\nano\meter} (purple) and \SI{679}{\nano\meter} (green).
    \textbf{b} We compute the expected dephasing of Rabi oscillations due to laser phase noise in a two-level approximation. 
    We approximate the effective three-photon phase noise power spectral density (PSD) with the trace of the \SI{688}{\nano\meter} laser (red trace). For comparison we also show the expected dephasing taking the \SI{679}{\nano\meter} phase noise PSD as input. Note that the relevant PSD for our three-photon coupling depends on the relative phase noise of the three participating lasers. Estimating the relative phase noise PSD requires further assumptions when the lasers exhibit similar noise spectra.}
    \label{fig:appendix_2}
\end{figure}
To estimate the effect of laser phase noise we analyze the in-loop beat signal of each laser with respect to the common frequency comb.
The beat signal is recorded with an electrical phase noise analyzer which provides the power spectral density, shown in \figref{fig:appendix_2}{a}, which has units of \si[per-mode=fraction]{\dbc\per\hertz}, in the following denoted by $\mathcal{L}_{\Phi}$. 
Since the \SI{688}{\nano\meter} laser exhibits considerably stronger phase noise than the other lasers, the relative phase noise of the three-photon couping is approximately
given by the noise of the \SI{688}{\nano\meter} laser alone. 
To estimate the effect of phase noise on our Rabi dephasing we compute the time-evolution of a resonantly driven two-level system described by the Hamiltonian
\begin{IEEEeqnarray*}{rCl}
    H &=& \frac{\Omega}{2} \ket{g}\bra{e} e^{-i\Phi(t)} + h.c.
\end{IEEEeqnarray*}
and average over multiple instances of the time-dependent phase $\Phi(t)$ which is characterized by the measured phase noise power spectral density. 
To generate samples of the time-dependent phase we start by interpolating the phase noise spectrum to an equidistant spacing $df$.
Samples of the phase traces are then given by
\begin{IEEEeqnarray}{rCl}
    \Phi(t) &=& \sum_f \sqrt{2\,S_{\Phi}  df} \cos\left(2\pi f t + \Phi_f\right)
\end{IEEEeqnarray}
where $S_{\Phi} = 2\cdot 10^{\mathcal{L}_{\phi}/10}$ and $\Phi_f \in [0,2\pi)$ is a randomly selected phase~\cite{Tsai2024,Rubiola2008}.
Averaging over $1000$ instances of sampled phase traces yields a Rabi dephasing in excellent agreement with the measurements indicating 
that phase noise is the main limitation (see \figref{fig:appendix_2}{b}).
%

%%%%%%%%%%%%%%%%%%%%%%%%%%%
%%% Atom-atom coherence %%%
%%%%%%%%%%%%%%%%%%%%%%%%%%%
\section{D. Atom-atom coherence}
To estimate the atomic coherence, we follow the analysis introduced in Ref.~\cite{Young2020}, which we briefly summarize here. 
A lower bound of the atomic coherence between states $\ket{g}$ and $\ket{e}$ is given by $\sqrt{\langle C_b \rangle /2} \leq |\rho_{eg}|$ with
\begin{IEEEeqnarray*}{rCl}
    C_b &=& \frac{1}{N(N-1)} \sum_{i \neq j} g_{ij}^{(2)}
\end{IEEEeqnarray*}
denoting the average of the $g_{ij}^{(2)} = \langle S_iS_j\rangle - \langle S_i \rangle \langle S_j \rangle$ correlator across all pairs of tweezers (with indices $i$ and $j$) within a given subset $b$. 
To compute the expectation values of the spin operators $S_i$ with eigenvalues of $-1$ $(+1)$ for the states $\ket{g}$ $(\ket{e})$ we average over all repetitions of the experiment and all phase values of the Ramsey-type sequence.
Next, we divide our tweezer array into subsets of $2 \times 2$ neighboring tweezer sites and average over all possible subsets to obtain the expectation value $\langle C_b \rangle$. The resulting lower bound for the atomic coherence is shown in Fig.~4b of the main text as a function of the Ramsey duration. 
We obtain an atom-atom coherence time of \atomatom{} from a Gaussian fit.

	%\bibliography{draft/manuscript/TripleMagic}
	\bibliography{TripleMagic}

\end{document}